\documentclass[acmsmall, nonacm]{acmart}
\usepackage{etoolbox}
\renewcommand\footnotetextcopyrightpermission[1]{}
\usepackage{tikz}
\usepackage{etoolbox}
\usepackage{pdfpages}
\usepackage{color,soul}
\AtBeginEnvironment{quote}{\singlespacing\small}
\usepackage{amsmath}

\makeatletter
\def\@copyrightspace{\relax}
\makeatother
\colorlet{change}{yellow!0}
\colorlet{keep}{black!0}

\newcommand{\rnr}[1]{{\sethlcolor{keep}\hl{#1}}}
\newcommand{\fed}[1]{{\sethlcolor{change}\hl{#1}}}

\begin{document}
\title{Social influence leads to the formation of diverse local trends}

 \author{Ziv Epstein}
 \email{zive@mit.edu}
 \author{Matthew Groh}
 \author{Abhimanyu Dubey}
 \author{Alex ``Sandy'' Pentland}
 \affiliation{%
   \institution{MIT Media Lab}
   \streetaddress{75 Amherst Street}
   \city{Cambridge}
   \state{MA}
   \postcode{02143}
\country{USA}
 }


\begin{abstract}
How does the visual design of digital platforms impact user behavior and the resulting environment? A body of work suggests that \rnr{introducing} social \rnr{signals} to content can increase both the inequality and unpredictability of its success, but has only been shown in the context of music listening.  To further examine the \fed{effect} of social influence on media popularity, we extend this research to the \rnr{context} of algorithmically-generated images by re-adapting Salganik et al's Music Lab experiment. On a digital platform where participants discover and curate AI-generated hybrid animals, we randomly assign both the knowledge of other participants' behavior and the visual presentation of the information. We successfully replicate the Music Lab's findings in the context of images, whereby social influence leads to an unpredictable winner-take-all market. However, we also find that social influence can lead to the emergence of local cultural trends that diverge from the status quo and are ultimately more diverse. We discuss the implications of these results for platform designers and animal conservation efforts. 

\end{abstract}


\keywords{platform design, social influence,  crowdsourcing}


\maketitle

\section{Introduction}
The explosion of information contained on modern online platforms requires users to use heuristics to both efficiently search through this information, and to make informed decisions. One such heuristic is the social signals of how other users \fed{have engaged} with the platform. \textit{Social influence} occurs when the decisions of a user are impacted by those of other users \cite{cialdini1998social}, and has been shown to be a key design dimension for contexts as varied as health behavior \cite{christakis2007spread}, political engagement \cite{bond201261} , collective behavior \cite{leskovec2010governance}, online book purchasing \cite{chen2008herd}, food ordering \cite{hou2017can}, and digital news engagement \cite{muchnik2013social, weninger2015random}. The ubiquity of social influence suggests how crucial a factor it is for platform designers seeking to jointly optimize for the quality and diversity of content online \cite{holtz2020engagement}. 

Perhaps the most influential study on how social influence and information hierarchy impact online platforms is that of \citet{salganik2006experimental}, informally dubbed the ``Music Lab'' experiment. In this study, the authors created an ``artificial cultural market,’’ where participants could listen to and download previously unknown songs. Critically, some participants were provided a layout which displayed information about previous participant's choices, while the others had no such knowledge. This experimental design allowed for causal identification of the role of social influence on both an individual's propensity to download songs, and on the dynamics of the ecosystem as a whole. In particular, Salganik et al found that \rnr{introducing} social influence increased the inequality of song success, as defined by the number of times they were downloaded. This suggests a \rnr{cascading} ``winner take all'' phenomenon, whereby social influence increased the availability of songs that were \rnr{perceived as} successful \rnr{by} past participants. They also found that social influence increased the \textit{unpredictability} of success, \rnr{as defined by the variation in a song's success across the worlds in an experimental condition}. \rnr{From these two findings, t}he authors infer that the underlying quality of a song only partly determines its final success: social influence causes a snowball effect that results in the emergence of local preferences. 

\rnr{The Music Lab experiment called to attention what is at stake when designing social influence into online platforms, but it remains unclear how to apply its conclusions to the design of modern social media platforms. For one, the original paper did not specify any mechanism or model to explain how social influence operates} \cite{krumme2012quantifying}. \rnr{Moreover, it is unclear how the findings in the context of music translate to other forms of media such as images.}

In the present paper, we show a conceptual replication of the original Music Lab study in a \rnr{context} that is \rnr{fundamentally} different from music --- the in silico evolution of AI-generated hybrid animals (which we call ``ganimals'') \cite{epstein2020interpolating}. In particular, we ask if similar patterns of results are observed in such a different context. This allows us to test the generalizability of the Music Lab to adjacent \rnr{contexts} (for a full characterization of the similarities and differences between the Music Lab and the present study, see Related Work). To do so, we built \textsc{Meet the Ganimals}, an online platform where users could generate and curate their own ganimals.  Previous work introduced this platform as a ``casual creator,’’ and evaluated how its random stimulus approach can efficiently search the possibility space of a GAN generator \cite{epstein2020interpolating}. Rather than focus on the usability of the platform itself, this work introduces an alternative \rnr{layout} to the ``Feed 'Em'' page of the system (see Figure \ref{fig:system_map}), and presents experimental results on the impact of such design interventions in the field. 

\rnr{As a collection of images of synthetically generated hybrid animals, ganimals represent a unique context to study social influence. Interpolations of image-based GANs are a new form of media about which, due to their unfamiliarity, most people have no preconceived notions.  In studies of social influence, previous knowledge of the content can introduce a key confound. Since ganimals are uniformly novel, both because of both the novelty of GAN-technologies, and also the vast possibility space of potential hybrid animals (see} \cite{epstein2020interpolating} \rnr{for a characterization of this possibility space), here we can study social influence excluding the potential confounding factor of previous experience (see Section 2.2 for a full discussion).  While the images as a whole are unfamiliar, the component parts --- bodies, faces, eyes, mouths, colors, backgrounds, positioning, etc. --- are well-studied affectively salient features} \cite{dydynski2018multisensory, genosko2005natures, goetschalckx2019ganalyze}. \rnr{As such, we hypothesize that the emotional valence that ganimals induce (falling in the uncanny valley of cute/creepiness) is highly \emph{subjective}, and therefore may be subject to social influence.

The use of ganimals also allows us to assess how the findings of the Music Lab might translate to the medium of images.  Relative to their text-based counterparts, image-based social media is on the rise (especially during the COVID-19 pandemic)} \cite{hu2014we, pittman2016social, masciantonio2021don}. \rnr{While images of ganimals of course differ from images on social media along many important axes, the process of rapidly searching through troves of emotionally salient and unfamiliar content and unconsciously deciding which to attend to (and engage with) may mirror some of the cognitive patterns of surfing social media} \cite{pennycook2021shifting, brady2020mad, vuilleumier2005brains, compton2003interface}.

A final key ingredient of the Ganimals platform is that it allows users to annotate the ganimals with morphological features. These rich ganimal-level covariates allow us to directly quantify the diversity and divergence of this online media ecosystem. 

This paper has five main contributions. First, we introduce the HCI community to the methods and results of the Music Lab study, and experiments that build on it. Second, we show a conceptual replication and generalization of the Music Lab to \rnr{the entirely different context of images}, which substantially increases the extent to which the HCI community can base theories and built systems on the original findings. Third, \rnr{we employ morphological embeddings to provide in-depth insight into both the diversity and divergence of digital ecosystems}. Fourth, we introduce a new visual display \rnr{layout}, called cloud view, which allows us to isolate the mechanistic features of ranked lists that drive the effects. Finally, we discuss how our findings and methodologies can be applied by systems designers to both quantify the emergent outcome of platform designs, and evaluate new visual \rnr{layout} designs.

\section{Related Work}
\subsection{Replications in Human Computer Interaction}
In response to the growing concern within the HCI community of prioritizing ``novelty'' over ``consolidation,'' the contribution of replications have been re-articulated \cite{greiffenhagen2013replication, wilson2014replichi, greenberg1991weak, hornbaek2014once, peng2011reproducible}. Recent HCI papers have replicated studies from online labor markets (like MTurk and Lucid) on topics such as misinformation \cite{epstein2020will}, visualization \cite{hu2019viznet, heer2010crowdsourcing}, input devices \cite{findlater2017differences}, and usable security \cite{redmiles2018asking}, but replications of large scale, virality based field experiments are more infrequent. By replicating the Music Lab, we show that such replications are well-scoped and useful for designing new systems.  

\subsection{Music Lab Experiments}
The Music Lab inspired a generation of experiments in artificial cultural markets designed to assess the impact of social influence and information design on collective behavior \cite{muchnik2013social, abeliuk2017taming, antenore2018songs, hogg2014disentangling, lerman2014leveraging, salganik2008leading}. Some have attempted to decouple social influence and item position, which were confounded in the original experiment \cite{hogg2014disentangling, lerman2014leveraging, abeliuk2017taming}. \citet{hogg2014disentangling} found that the impact of position is twice that of social influence itself. \citet{abeliuk2017taming} found that ranking \rnr{positively} affects unpredictability more than social influence does, and that combining ranking by quality and social influence allows high quality stories to become ``blockbusters.''

 In a follow-up paper to the Music Lab, \citet{salganik2008leading} seeded songs with false and arbitrary initial signals of \fed{popularity}, and looked at how those initial signals affected the market equilibrium. They found that while certain songs had a ``self-fulfilling prophecy,'' the best songs were able to recover their popularity in the long run. They also found that the initial distortion of the market information reduced correlations between appeal and popularity, and led to fewer overall downloads. Building on this work, \citet{shulman2016predictability} found that while it is hard to predict an item's final popularity, ``peaking'' at early adopters provides a highly effective framework for predicting future success. 

\citet{antenore2018songs} conducted a Music Lab experiment where they found no evidence of an effect of social influence. Critically, their experiment only contained 10 songs to ``avoid as much as possible the interference attributable to choice overload.'' The  differences they observed are due to the fact that the small number of songs meant every participant could try each and every song, and did not need to use social signals as a heuristic to ``avoid the high cognitive cost of exploration.'' Yet this heuristic is a critical feature of social influence, since most markets of interest involve too many items to try every one (see Section 2.3, below). In addition, their experiment was not a web-based study and occurred in a computer lab under the direct supervision of the experimenter. This induced a focused mindset divorced from the actual cognitive context where most cultural markets occur (e.g. where people are distracted, overwhelmed and must rely on heuristics).  

In contrast, like the Music Lab, our study recruited subjects from the internet via their intrinsic interest in the subject matter (not a financial incentive). We also took precautions to ensure the content participants saw would be unfamiliar to them (since previous knowledge would introduce a confound). The original Music Lab experiment went to great lengths to ensure the songs selected were unknown to subjects, such as restricting to bands that played in less than 10 states, played less than 15 concerts in the past 30 days, had less than 30k hits on their PureVolume page, and had not played at the Warped Tour. The authors themselves admit that their restriction criteria are ``ultimately arbitrary.'' By focusing on ganimals, we could be sure that the stimulii were inherently and uniformly unfamiliar without relying on ad hoc restrictions like the Music Lab \fed{did}.

\rnr{Our conceptual replication of the Music Lab has several key differences from the original study that are important to highlight. First and foremost, we focus on the domain of (AI-generated) images, rather than music. A critical theoretical difference between these mediums is that in the original Music Lab, participants first decided whether they wanted to listen to song based on social and other signals, whereas in the image context the impression of the media is immediate and inherently entangled with other signals (see Section 2.3 for a full characterization of this two-stage process, and how the image context relates). A second difference is we use user-annotated ganimal morphology to directly assess the divergence and diversity of the media environment. Finally, we also manipulate the visual display layout in order to isolate the components of ranked lists that drive the effects. }

\subsection{A theoretical model for cultural markets}
\citet{krumme2012quantifying} observe that a market for songs involves a two-stage process: the participant first chooses which song they will listen to, and then after listening, decides whether or not to download that song. In the first stage, the only information the participant has to decide if they will listen to the song is the name of the song and band, and \fed{also} the social signals if they are in the social influence condition.  \citet{krumme2012quantifying} found that social influence is only present in the first stage, and that the probability a user downloads a song is conditionally independent to if they clicked on it \cite{krumme2012quantifying}. This two-stage model explains the findings of \citet{antenore2018songs}: with ample time and only 10 songs, participants were able to ``try'' each and every song and thus social influence did not factor into the ``buying'' stage. \citet{abeliuk2017taming} use this formulation to derive a metric for quality --- the conditional probability of downloading a song given that it was sampled --- that they recommend for optimizing. 

\rnr{This ``try and buy'' model of cultural markets has also been mathematically characterized as a individual-level heuristic that results in collective Bayesian rationality} \cite{krafft2021bayesian}. \rnr{In particular, individual agents locally utilizing a ``try and buy'' heuristic corresponds to a regret-minimizing solution to a population-level exploration-exploitation dilemma (e.g. Thompson sampling). }

In the context of ganimals, where the content are images, we study \textit{only the first stage} of this two-stage process. It is appropriate to compare this first stage of the process to the Music Lab, since it is in this stage (not the second) where social influence is present.  



\subsection{Quantifying diversity and divergence}
A key aspect of social influence is its capacity to decrease the diversity of an information environment \cite{lorenz2011social, gillani2018me, sunstein1999law,yardi2010dynamic}. \citet{sunstein2001republic} has argued that social media has  engendered a polarized culture where people do not seek out new information. The salience of group identity online can impact the perceived and actual diversity of the resulting ecosystem by fostering an in-group/out-group mentality \cite{brady2020mad,  yardi2010dynamic}. 

Recent work has shed light on design features that moderate the relationship between social influence and diversity. In the context of music listening, \citet{holtz2020engagement} found that personalized recommendation decreased diversity within users, but increased diversity across users. \citet{pescetelli2020diversity} found that \rnr{increased} diversity increases collective intelligence for large groups ($\approx25$ people) but decreases it for small groups ($\approx5$ people). \citet{lorenz2011social} found that social influence undermines the wisdom of the crowd by mitigating the diversity of the crowd's responses without improving \fed{upon} its collective errors.  


However, the impact of social influence on diversity is understudied in the Music Lab context. This is because the content in those artificial cultural markets does not typically include any covariates, so the authors focus their analyses only on engagement metadata, such as popularity. Ganimals, however, are annotated with their morphological characteristics, which in turn allows us to assess how the experimental conditions affect the distribution of these characteristics across worlds. 

\subsection{Information design for social influence}
\rnr{A growing body of work within HCI has explored how information design can impact how users interact with systems} \cite{klemmer2000designer, doosti2017deep, dong2012social,wexelblat1999footprints, introne2012design}. \citet{toth1994effects} \rnr{explored the modality of feedback in a group discussion paradigm, and found that 2D graphics can augment normative and inhibit informational social influence. } \citet{hullman2011impact} \rnr{showed that social signals affected graphical perception accuracy in a linear association task. They also demonstrate a cascade pattern, such that initial inaccurate guesses can erroneously affect the responses of subsequent participants.  Romero found a substantial effect of early respondents in Doodle polls} \cite{romero2017influence}, \rnr{whereby the first few respondents of a poll can dramatically influence the behaviors of subsequent respondents.}  In the context of online gift giving, \citet{kizilcec2018social} found that receiving a gift causes individuals to give more gifts in the future, and that designing observability into a system made gift-giving more socially acceptable \cite{kizilcec2018social}. \citet{sharma2016distinguishing} \rnr{introduced a statistical procedure for distinguishing between personal preferences and social imitation behavior, and find that a large  majority of user actions reflect personal preference rather than copy-influence on a music recommendation website.} \citet{wijenayake2020quantifying} \rnr{investigated how design features like user representation, interactivity, and response visibility impact conformity. They find not only main effects in differences in group size, task objectivity, and perceived self-confidence, but also interactions between interactivity and response visibility.}

The Music Lab and its follow-ups have primarily focused on linear lists and grids of music \cite{salganik2006experimental,salganik2008leading,antenore2018songs} and scientific articles \cite{abeliuk2017taming}. This standard display \rnr{layout}, also employed by social media platforms as a ``newsfeed,'' involves scrolling through large lists of content.  We maintain the use of the list, but also introduce a new visual display, inspired by tag clouds and the designers outpost \cite{klemmer2000designer}, called ``cloud view.'' In contrast to the Music Lab’s two experimental conditions (independent and social influence), we cross those conditions with showing the participant either the standard ranked list view, or the alternative cloud view (see Section 3.1 for more details on the experimental design). By experimentally varying social influence and the type of \rnr{layout}, we can decouple the relative effects of each, which serves two critical purposes. First, it allows us to separate screen location from popularity information, which were confounded in the original Music Lab experiment (that is, for the social influence condition, the more popular items where both higher in the list, and designated by their popularity --- here, the ecosystem view allows us to disentangle these two factors). Second, it allows us to see what results are dependent on having to scroll through all the ganimals individually, versus all of them being presented together.

\section{Methods}
\begin{figure}[h]
 \includegraphics[width=0.99\textwidth]{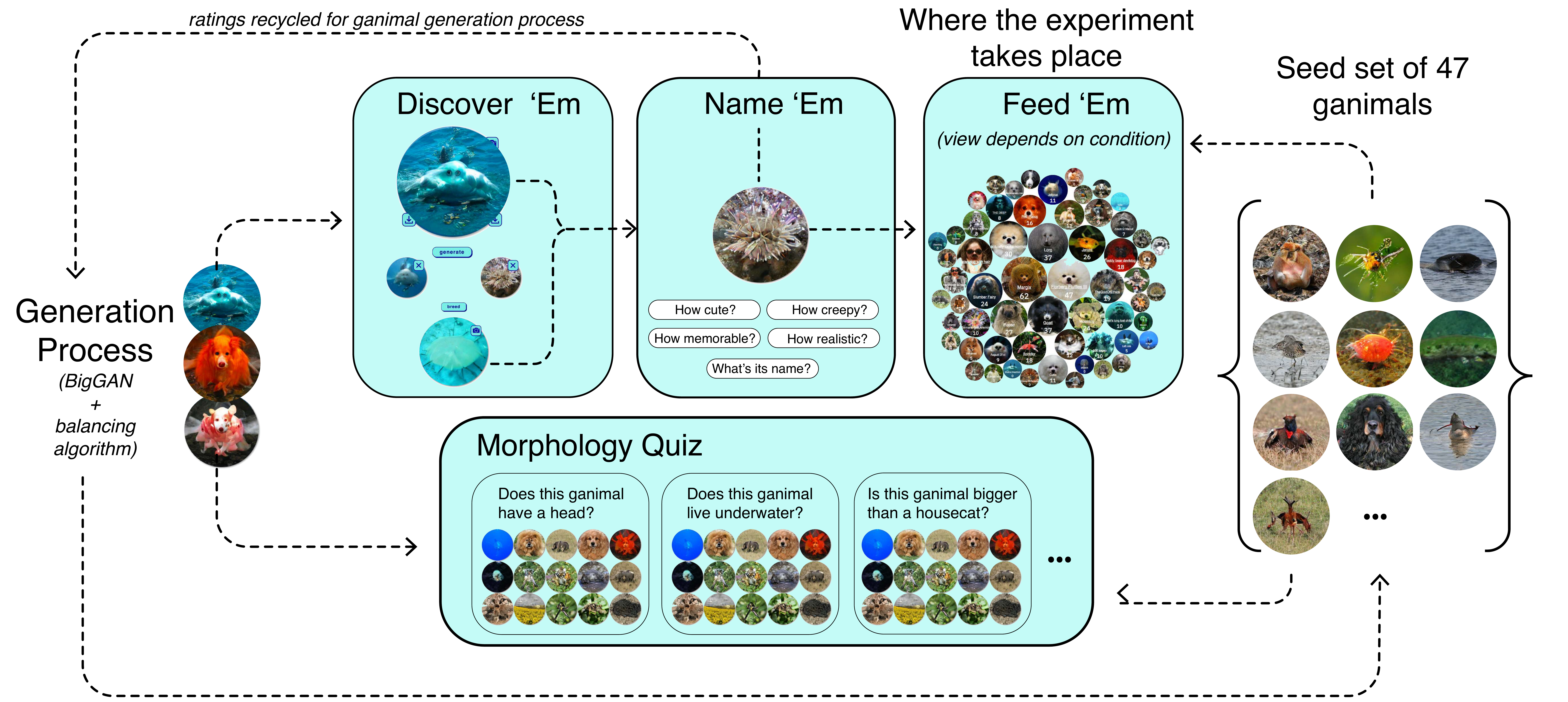}
 \caption{Schematic of the \textsc{Meet the Ganimals} architecture. Ganimals are generated via BigGAN and a process that balances exploring new and existing ganimals. Participants can discover and breed ganimals in the Discover 'Em page, and name and rate their favorites in the Name 'Em page. The experiment took place in the Feed 'Em page, which also includes 47 seed ganimals. Participants can also characterize ganimals in the Morphology Quiz. Adapted from \cite{epstein2020interpolating} with permission.}
  \label{fig:system_map}
\end{figure}

\textsc{Meet the Ganimals} is an online platform where individuals can generate and curate ``ganimals'' - AI-generated hybrid animals \cite{epstein2020interpolating}. A schematic for the system is shown in Figure~\ref{fig:system_map}. Ganimals are generated by blending animal categories in BigGAN  \cite{brock2018large} in a way that balances exploring new hybrids, and exploiting existing signals for ganimal quality (see \cite{epstein2020interpolating} for a characterization of this algorithm). In the Discover 'Em page, participants could interact with the generated ganimals and breed their own. Once they found a ganimal they like, they could ``discover'' it by naming their ganimal and rating how cute/creepy/realistic/memorable it is on the Name 'Em page.\footnote{these subjective signals are then recycled for future ganimal generation} Discovered ganimals appeared in the Feed 'Em page, where users could ``feed'' (i.e. cast a vote for) the ganimals they liked the best. Separately, in the Catalogue 'Em page, participants could rate the morphological traits of the ganimals (see Section 3.3 for more details). \rnr{Screenshots of and more information about all of the pages can be found in the Supplementary Materials. }
\subsection{Experimental Design}
The experiment itself took place in the Feed 'Em page. As participants arrived to the platform, they were randomly assigned to one of four conditions (independent list, independent cloud, social influence list, social influence cloud) distinguished solely by the availability of information about the prior decisions of others, and the visual display of the ganimals in the Feed 'Em page.

\begin{figure}[h]
\centering
 \includegraphics[width=0.99\textwidth]{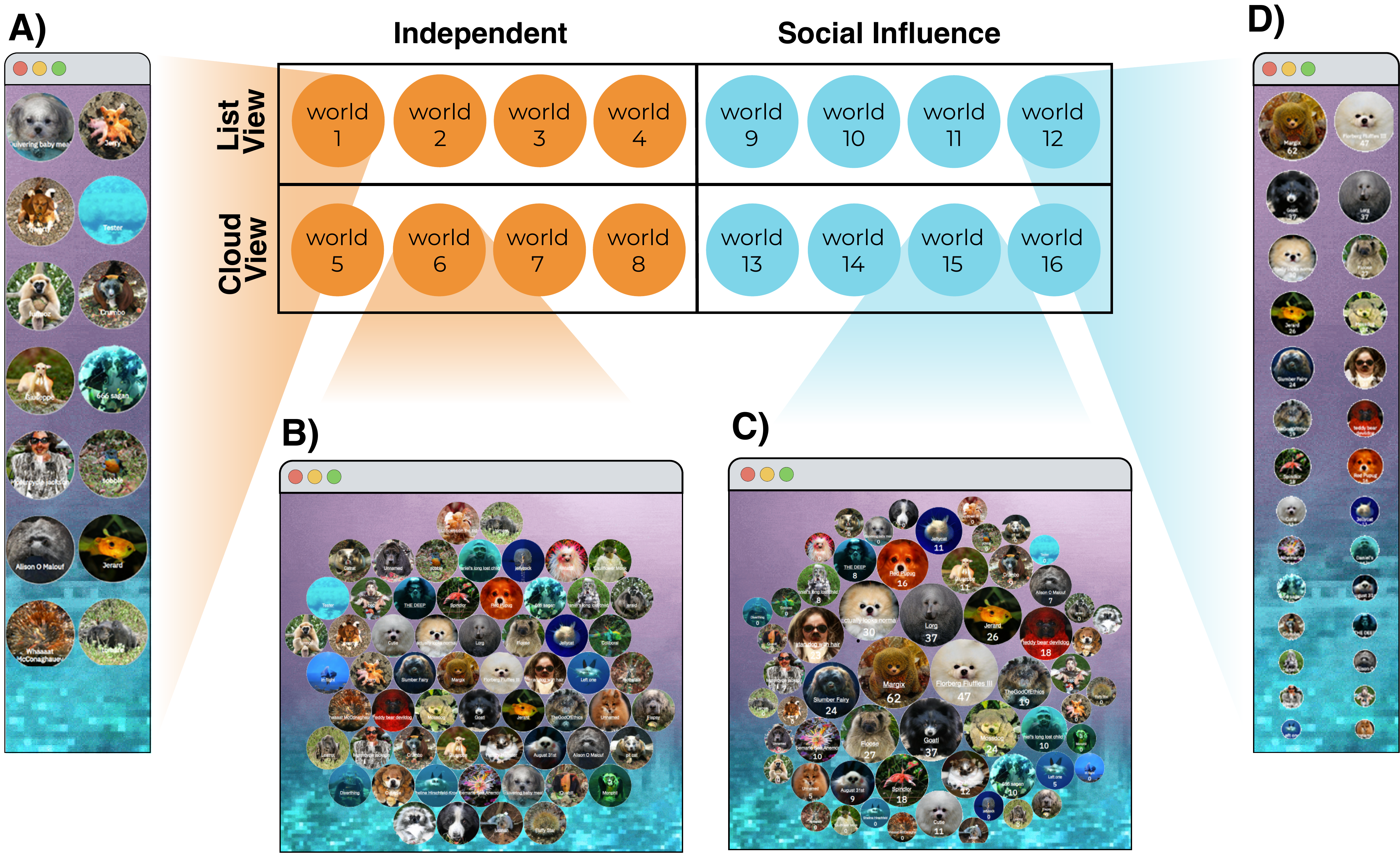}
 \caption{Overview of our experimental design, with (stylized) screenshots of the Feed 'Em page in the four conditions designated by the letters: A) independent list, B) independent cloud, C) social influence cloud, D) social influence list.  There are four worlds for each the experimental conditions.}
  \label{fig:exde}
\end{figure}

In the independent conditions, each ganimal was displayed at the same size, whereas in the social influence conditions, the ganimal's size was proportional to the number of votes it had from previous participants, and displayed this number as well as its name (see Figure~\ref{fig:exde}). Therefore, participants in the social influence conditions were provided a signal on the preferences of past participants, which they could use to make their own decisions.  All users could ``feed'' (i.e. cast a vote \fed{for}) the ganimals they liked the best, and the interface included the instructions: ``Ganimals need food to survive. To feed a ganimal, click on its image. To learn more about that ganimal, click on its name.''  For both the independent and social influence conditions, whenever a participant voted on a ganimal, it grew a bit larger. In the list conditions, ganimals are rank ordered and displayed by number of votes in a grid which has two columns in desktop view and one column in mobile view. In the cloud view, ganimals are displayed in a spatial circle pack, with larger ganimals often (but not always) in the center.  

Within each of the four experimental conditions, participants were randomly assigned to one of four ``worlds'' (for a total of 16 --- see Figure~\ref{fig:exde}), each of which \fed{evolved} independently of the other fifteen. In particular, participants only saw ganimals discovered and votes cast by others in their world, and the ranking (for list view) and visualization (for cloud view) of ganimals was based only on votes in that world. \rnr{ A randomly chosen ``seed set'' of 47 ganimals was used to initialize each of the sixteen worlds (such that all worlds started with the same set of initial ganimals, see Section 1 of the Supplementary Materials for more information).}

Irrespective of condition, participants could choose between 60 different ganimals to feed. At any given time, this set of 60 included 30 of the top voted ganimals in that world, and 30 of the most recently discovered ganimals in that world. The top rated ganimals were initialized with the seed set, but at most 30 of the them were shown to the user, and that number was much less once participants in a given world started voting (i.e. no ganimals with zero votes were displayed after thirty ganimals were voted on in that world). 
\subsection{Recruitment, reliability and robustness}

From April 26th to June 26th 2020, 44,791 ganimals were generated, 8,547  ganimals were bred, and 743 ganimals were named by a total of 10,657 users. In the Feed 'Em page 2,370 votes were placed on 434 ganimals by 549 users. Of these 549 users, 18\% were on mobile, while 81\% where on desktop, and they predominately hailed from Russia (27\%), USA (23\%), Ukraine (16\%), and Japan (3\%). We did not collect other demographic information. 

Participants were recruited through word of mouth and social media, bolstered by a \fed{climate fiction (cli-fi) world-building campaign} (learn more here: \url{https://www.youtube.com/watch?v=I-Fc4nQK_5Q}).

A critical part of our experiment was making sure there was no information contamination between worlds and conditions. We used cookies to ensure that each user would be placed in the same world if they returned to the website at a later time. To prevent contamination from exposure to other ganimals after they first experience the website \fed{(e.g. from elsewhere online)}, we only counted votes that occurred within 2 hours of the participants first visiting the site. To mitigate the impact of a few power users, we only counted the first 10 votes of each participant, as recorded through cookies.

We preregistered our primary hypotheses, primary analyses and sample size, which are available at \url{https://aspredicted.org/65nv7.pdf}. \footnote{\rnr{Pre-registration is a framework that allows researchers to specify which analyses are \textit{a priori} and which are \textit{post hoc}. This strengthens the validity of statistical analyses, and can mitigate publication bias} \cite{nosek2019preregistration}.}
Because the experiment received less media attention than anticipated due to Covid-19, we deviated from our preregistration \fed{in} several ways. First, we stopped data collection early, but did not look at any results before doing so (due to the number of monthly active users, we would never reach 1K people --- the number specified by our pre-registration). Second, this smaller $N$ meant we were unable to block on cohort, since there was only a single cohort with its own set of worlds. As such, for all our analyses, we report results only for the first cohort, and also do not look at the interaction between treatment arms, as we are underpowered to do so with only 16 observations. Finally, the limited number of participants also meant that computing a joint distribution between popularity and a high-dimensional feature embedding for the diversity measure was under-specified. Therefore, we instead directly compute the entropy of the morphological traits, which is ultimately a more interpretable metric.

This study was approved by the MIT COUHES committee.


\subsection{Crowd annotation of ganimal morphology}
Participants could also provide information about the morphology of ganimals. We worked with a professional zoologist to assemble 10 traits that would characterize the variability in possible ganimals (for the full list of traits, see Table~\ref{tab:morph_feat}). Within the ``Morphology Quiz'' section, the user was provided with 16 ganimals for each morphological trait, and was asked to select the ones that exhibited that trait. 
\begin{table*}[h]
\centering
  \caption{Morphological traits used.}
  \label{tab:morph_feat}
  \begin{tabular}{|ll|ll|}
    \hline
     Trait name & Question Asked&Trait name & Question Asked  \\
    \hline\hline
    Head & Does this ganimal have a head? & Size & Is this ganimal bigger than a house cat?\\ 
    Eyes & Does this ganimal have eyes?& Underwater & Does this ganimal live underwater?  \\ 
    Mouth & Does this ganimal have a mouth? & Feathers & Does this ganimal have feathers? \\ 
    Nose & Does this ganimal have a nose? & Scales & Does this ganimal have scales?  \\ 
    Legs & Does this ganimal have legs?& Hair & Does this ganimal have hair? \\
    \hline
  \end{tabular}
\end{table*}
For each ganimal, we computed the average response for each morphological trait across responses. With a 1 coded as exhibiting the trait and 0 coded as not, the average response represents the likelihood or extent to which a given ganimal exhibits a given trait.  14,348 ratings were provided for 1,250 ganimals by 177 users of the \textsc{Meet the Ganimals} platform (48 of these raters were also participants in the experiment). Many of these 1,250 ganimals did not appear in the actual experiment, so we restrict our attention to the 449 ganimals that received at least one vote. 

21\% of the 449 ganimals \rnr{were missing at least one morphological trait (due to limits in data labeling)}, and among this subset, there was an average of 6.13/10 \rnr{non-missing} traits. Furthermore, multiple users often rated the same trait of a given ganimal: among this subset, an average of 2.4 users rated each non-missing trait. We find no statistical difference in the number of \rnr{missing} traits across conditions ($p=0.351$ for social influence, $p=0.175$ for information design).



\section{Results}
We investigate the role of social influence and information design on five outcomes: inequality, unpredictability, diversity, divergence and engagement. As a replication of \citet{salganik2006experimental}, we find evidence that social influence increases inequality and unpredictability. We also find evidence that social influence increases the morphological divergence and diversity of worlds. For each set of analyses, we focus on the list view results since those are directly comparable to the literature, but also show the cloud view results for contrast. 

\subsection{Inequality}
\begin{figure}[h]
 \includegraphics[width=0.99\textwidth]{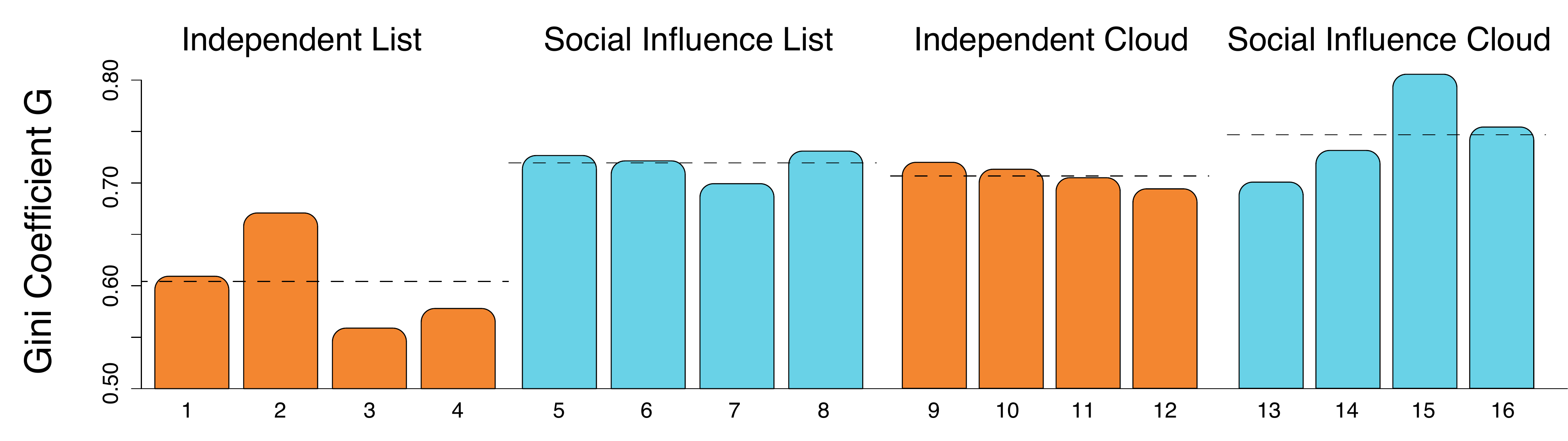}
 \caption{Inequality of success for independent list (orange left), social influence list (blue left), independent cloud (orange right) and social influence cloud (blue right) worlds, with number corresponding to world. Dashed line corresponds to the average Gini coefficient of all worlds within a given condition.}
 \label{fig:gini}
\end{figure}

\rnr{To make our results comparable to} \citet{salganik2006experimental}, we also use the Gini Coefficient \cite{bendel1989comparison} to assess the concentration of the votes of the ganimals in each world.
Figure~\ref{fig:gini} and Table~\ref{tab:gini} show the effects of social influence and \rnr{layout} on inequality, as measured by the Gini coefficient. For both the list and cloud displays, we find that worlds with social influence exhibit more inequality than the worlds where participants made independent decisions ($p=0.002$, \fed{preregistered}), which is a direct replication of the Music Lab experiment.
\begin{table*}
\centering
  \caption{World-level linear regression predicting inequality. }
  \label{tab:gini}
  \begin{tabular}{lcccl}
    \toprule
     &Estimate & Standard Error & $t$-value & $p$-value \\
    \midrule
    {Social Influence} &   0.077& 0.019 &3.981&0.002  \\
    {Cloud} & 0.066& 0.019& 3.379&0.002 \\
    {Intercept} & 0.623&  0.016& 36.9&<0.001 \\
    \midrule
          {Adjusted $r^2=0.628 $, N=16} &  & & & \\
    \bottomrule
  \end{tabular}
\end{table*}

As a preregistered robustness check, we use Fisherian randomization inference (FRI) to compute an exact $p$-value \cite{imbens2015causal}. Fisherian randomization inference is a non-parametric approach to computing p-values that does not require modeling assumptions about potential outcomes. To perform FRI, we create 10,000 permutations of the assigned world treatments, and recompute the t-statistic for each. We then compute a p-value by assessing the fraction of permutations that yielded t-statistics larger than the t-statistic observed in the actual data. We find that $p_{\text{fisher}}<0.001$. 

As a posthoc robustness check, we perform bootstrapping at the world level and count the fraction of bootstrap samples with mean Gini greater for social influence than independent. We find that in 100\% of the 10,000 bootstrap samples, the mean Gini is greater for social influence worlds than independent worlds. 

We also find a main effect of design, whereby worlds with the cloud display exhibited more inequality than the worlds with the list display ($p=0.005$ for regression and $p_{\text{fisher}}=0.0015$, preregistered, and the mean Gini is greater for cloud view worlds than list view worlds for 100\% of the 10,000 bootstrap samples).

When restricting only to list view worlds, we find that social influence is significantly associated with inequality ($p= 0.004$, posthoc). This suggests a stronger association between social influence and inequality in the list view, where participants rely more heavily on social signals instead of scrolling through all 60 ganimals (versus cloud view where all are visible and thus social influence is a less important cue).

\rnr{As additional robustness checks, we reran the main analyses restricting only to ganimals with one or more vote, and using additional measures of concentration (following the robustness checks of} \citet{salganik2006experimental}) \rnr{, which are reported in Section 3.1 of the Supplementary Materials. The results are similar across model specifications, except for layout design, where the effect of Cloud View is not significant  for the Herfindahl index model (this may be because, unlike Gini or the coefficient of variation, the Herfindahl index is correlated with the number of ganimals in a given world.)}

\subsection{Unpredictability}
To measure the unpredictability of each condition, we follow the Music Lab and compute the average  difference  in  market share  for  that  ganimal  between pairs of realizations. The one critical difference in our setup is, unlike the Music Lab, not all ganimals necessarily appear in all worlds. Thus, we only consider pairs of worlds \textit{for which that ganimal actually appeared}. In particular, we first compute the market share of votes $m_i^{(j)} = v_i^{(j)} / \sum_\ell v_\ell^{(j)}$ for each ganimal $i$ in each world $j$. Then, we compute the average difference in market share for all pairs of worlds $j$ and $k$  within that condition $S$, and then average across ganimals to get a unpredictability score $u$ for each pair of worlds:
\begin{align*}
    &u_{j,k} =\frac{1}{N} \sum_i \sum_{j \in S} \sum_{k \in S: j \neq k} \left|m_i^{(j)}-m_i^{(k)}\right|.
\end{align*}

This gives us $\binom{4}{2}=6$ unpredictability scores per condition.  \rnr{To reduce noise in our unpredictability estimates, we deviated from our preregistration plan in two ways. First, because the vast majority of engagment was with ganimals not in the seed set (71\%), we considered all ganimals, not just seed ganimals. Second, since many ganimals where not seen and thus could not even be voted on, we restricted only to ganimals with atleast one vote (see SI Section 3.2 for full justification for these changes, as well as additional analyses for our original measure).}

\begin{figure}[h]
 \includegraphics[width=0.3\textwidth]{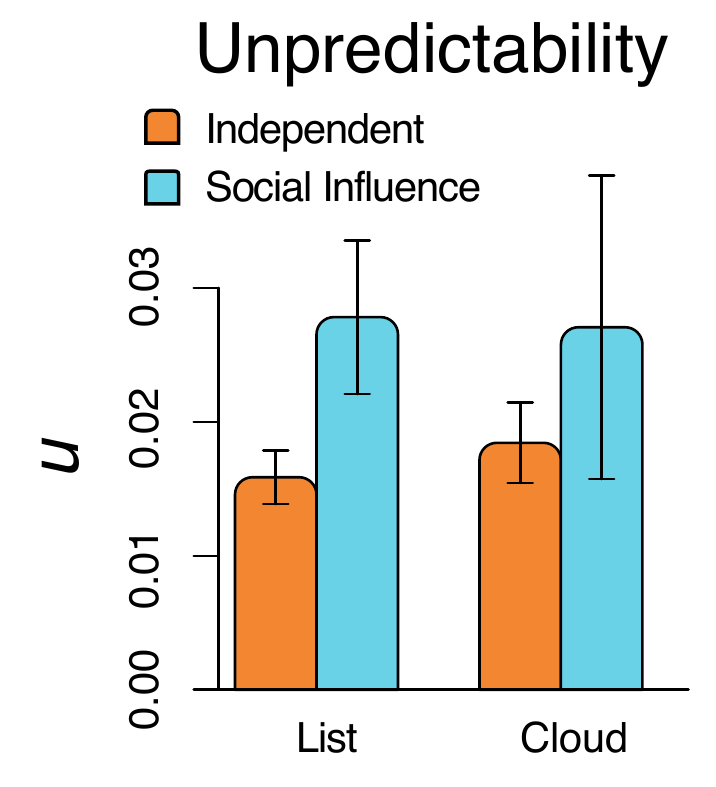}
 \caption{Unpredictability of success for the four conditions, computing using all ganimals with more than one vote.}
 \label{fig:unpred}
\end{figure}

\begin{table*}
\centering
  \caption{Linear regression predicting unpredictability. }
  \label{tab:three}
  \begin{tabular}{lcccl}
    \toprule
     &Estimate & Standard Error & $t$-value & $p$-value \\
    \midrule
    {Social Influence} &   0.010& 0.003 &3.112&0.005  \\
    {Cloud} & 0.001& 0.003& 0.275&0.786 \\
    {Intercept} & 0.016&  0.002& 5.831&<0.001 \\
    \midrule
          {Adjusted $r^2=0.252 $, N=24} &  & & & \\
    \bottomrule
  \end{tabular}
\end{table*}

The left panel of Figure~\ref{fig:unpred} and Table~\ref{tab:three} show the effects of social influence and \rnr{layout} on unpredictability, as measured by the average unpredictability across world pairs.  We find that worlds with social influence exhibit significantly more unpredictability than the worlds where participants made independent decisions ($p=0.005$, posthoc), but found no effect of the cloud \rnr{layout} ($p=0.786$).  When restricting only to list view worlds, we find a significant association between social influence and unpredictability for all ganimals with more than one vote ($p=0.003$). This suggests a stronger association between social influence and unpredictability in the list view, where instead of browsing all ganimals, participants relied more on social signals.

\subsection{Divergence and Diversity}
The 48 songs from the original Music Lab experiment did not include a rich set of covariates, so the authors focused their analyses only on the songs' popularity. The \textsc{Meet the Ganimals} platform, in contrast, allows users to annotate ganimals across 10 morphological traits (see Figure~\ref{fig:div2plots} \rnr{for the morphology of four exemplary ganimals, and} Table~\ref{tab:morph_feat} for more details). These ganimal-level covariates allows us to characterize and compare the kind of ganimals that evolved across worlds. In particular, we focus on the divergence of these features between worlds and the diversity of these features within worlds. 

\begin{figure}[h]
 \includegraphics[width=0.7\textwidth]{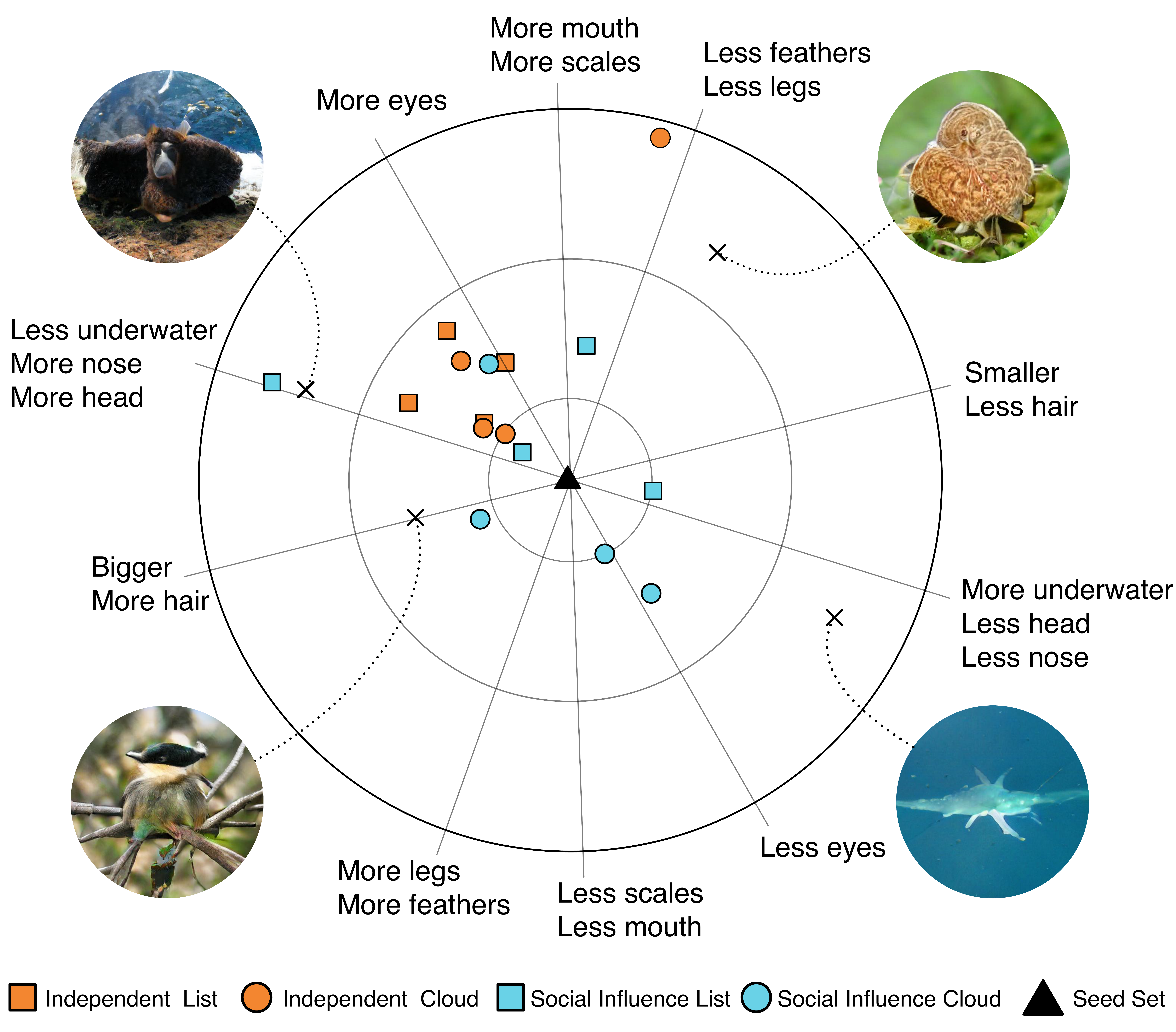}
 \caption{Morphological embeddings of the 16 worlds, relative to the morphology of the seed set of ganimals. Lines correspond to orthonormal projections for each of the 10 morphological features. The morphological embeddings of four exemplary ganimals is also shown.}
 \label{fig:divergence}
\end{figure}

\begin{figure}[h]
 \includegraphics[width=0.99\textwidth]{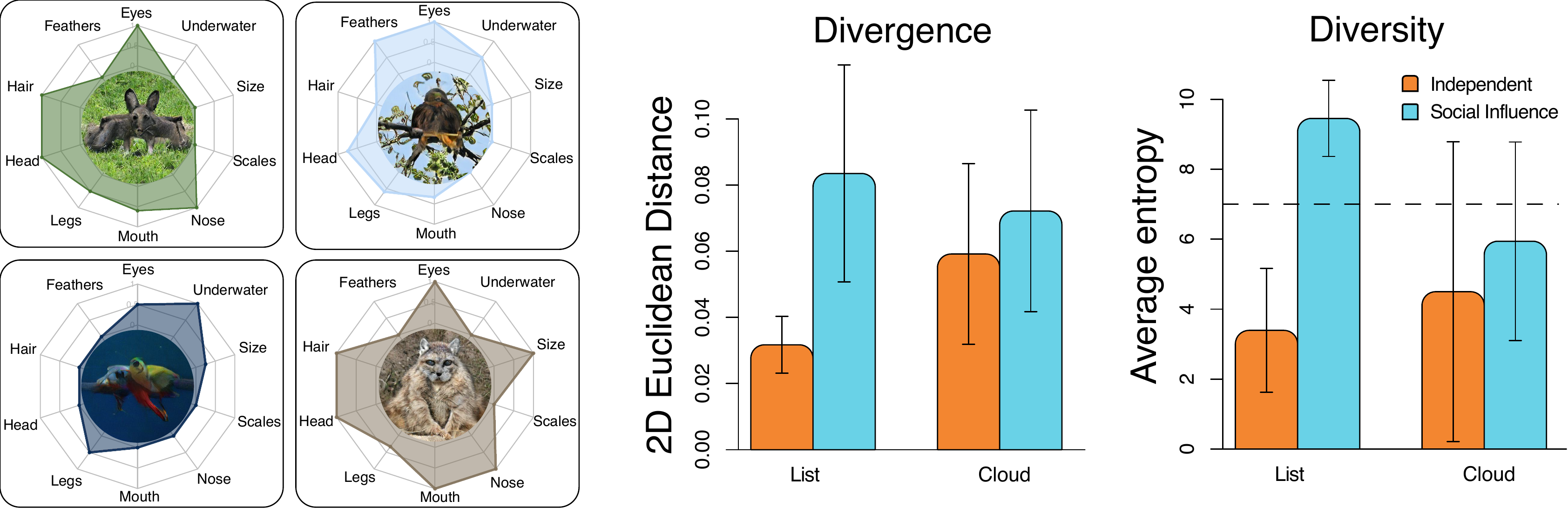}
 \caption{Left: Morphological features of four ganimals: a alligator/basenji hybrid (green), a kite/spider monkey hybrid (light blue), a hammerhead shark/macaw hybrid (dark blue) and a lynx/meerkat hybrid (brown). Center: Divergence across conditions, measured using 2-D Euclidean distance. Right: Diversity across conditions, measured by entropy of the multivariate gaussian fit to each world, with the dashed line corresponding to the imputed entropy of the seed set.}
 \label{fig:div2plots}
\end{figure}

For each world, we fit a multivariate gaussian with mean $\mu_j$ and covariance $\Sigma_j$ to the feature vectors of each ganimal in that world with one or more vote. Thus $\mu_j$ represents the average morphology of world $j$ (e.g. its local trend). We use principal component analysis (PCA) to collapse these world-level average feature embeddings $\mu_j$ into a 2D space for visualization (offset such that the average feature embeddings of the seed ganimals corresponds to the origin). In this 2D space, the PCA parameters define orthonormal lines for each of the 10 morphological features. The map of the 16 worlds in this coordinate space, as well as exemplary ganimals, are shown in Figure~\ref{fig:divergence}. As shown, the majority of worlds centered around ganimals with eyes, a nose, a head, and that do not live underwater. Several worlds (all social influence) diverged and are centered around underwater ganimals without eyes, nose or a head. 

To quantify these visual intuitions, we measure the Euclidean distance between the 2-D points in Figure~\ref{fig:divergence}, and then compute the average distance between pairs of worlds within a given condition.  These results are shown in Table~\ref{tab:div}. In the additive regression model, we find a significant effect of social influence on divergence ($p=0.030$). We find no effect for display type ($p=0.568$). 

\begin{table*}
\centering
  \caption{Linear regressions predicting diversity and divergence. $^{.}$ refers to $p\leq0.1$, $^*$ refers to $p\leq.05$, $^{**}$ refers to $p\leq.01$, $^{***}$ refers to $p\leq.001$. The value in parentheses under each coefficient is the standard error.}
  \label{tab:div}
  \begin{tabular}{lllll}
    \toprule
     Outcome variable (DV)&   Divergence (2D Euclidean) &Diversity \\
    \midrule
    {Social Influence} &0.032$^{*}$  &  3.75$^{**}$   \\
                        &(0.014) & (1.50)\\
    {Cloud} &0.008 & -1.20\\
                &(0.014)   & (1.50) \\
    {Intercept}  & 0.041 $^{***}$&4.55$^{***}$  \\
              &(0.012)  & (1.32) \\
    \midrule
    N   &24& 16 \\
      \midrule
          {Adjusted $R^2$} &0.014&  0.245\\
    \bottomrule
  \end{tabular}
\end{table*}

\rnr{As additional robustness checks, we quantify divergence using two additional measures, 10-D euclidean distance and Fr\'{e}chet distance, which are reported in Section 3.3 of the Supplemental Materials. We find a marginal ($p_{10D}=0.053$) and no effect ($p_{frechet}=0.607$) of social influence, respectively, and a marginal effect of display type in both cases ($p_{10D}=0.095$ and $p_{frechet}=0.053$).}

However, when restricting only to list worlds, we find a significant effect of social influence on divergence for all three measures ($p_{2D} =0.013, p_{10D}=0.021, p_{frechet}=0.021$). This suggests that in the list view, where scrolling through all the ganimals is cumbersome, social influence is an important cue. However in the cloud view, all the ganimals are easily available via a quick scan, so those worlds have high divergence regardless of social influence. 
%

These results suggest that there is dramatic variation in morphology across worlds. But how does that compare to the variation in morphology within worlds? To answer that question, we calculate and compare the morphological diversities of each world.  To compute the diversity of a given world $j$, we again use the multivariate gaussians with mean $\mu_j$ and covariance $\Sigma_j$ fitted to the feature vectors of each ganimal in that world with one or more vote. Then, we calculate the entropy of that distribution:
\begin{equation*}
h_j = \frac{1}{2} \log\det (\pi e \Sigma_j)). \tag{diversity}
\end{equation*}

The average morphological diversity of the four conditions is shown in Figure~\ref{fig:div2plots}. We find that worlds with social influence are more diverse than worlds where participants made independent decisions ($p=0.02208$,  $p_{\text{fisher}}=0.0099$). We also bootstrap the diversity at the world level, and find that the mean diversity is greater for social influence worlds than independent worlds for 99.991\% of the 10,000 bootstrap samples. We find no difference in morphological diversity between the list and cloud views for the main regression ($p=0.3842$), FRI ($p_{\text{fisher}}= 0.8161$), or through bootstrapping (14.51\% of bootstrap samples have more diversity in cloud view conditions than list view).

\subsection{Engagement}
\begin{figure}[h]
 \includegraphics[width=0.99\textwidth]{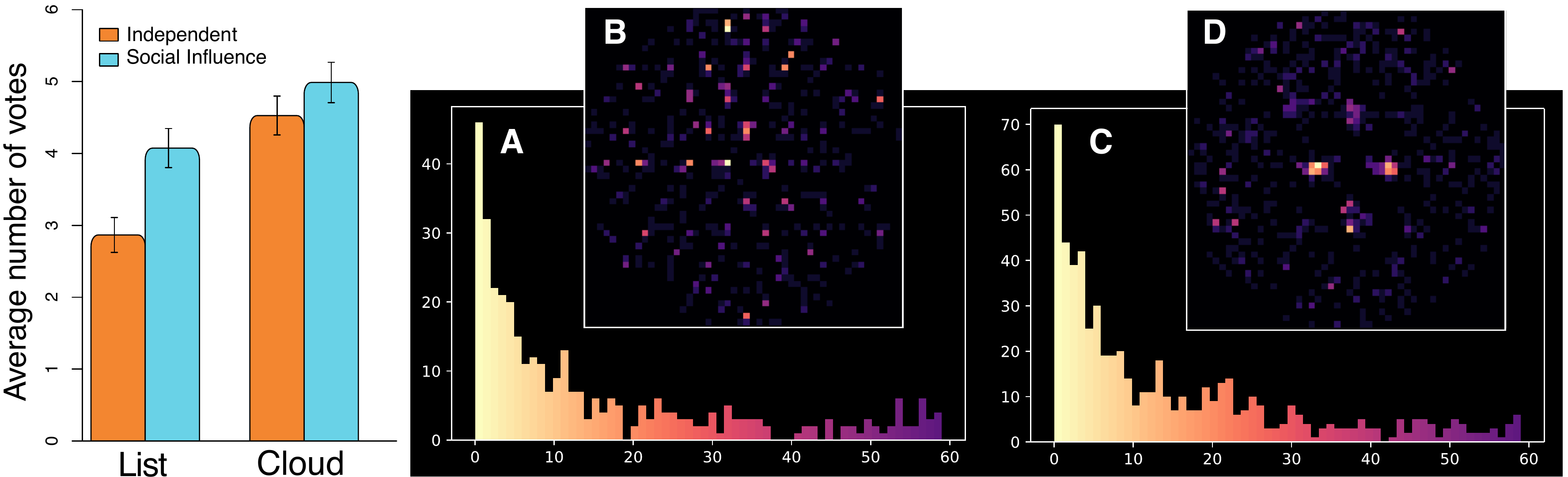}
 \caption{Left: Number of votes participants casts across conditions. Right:  Distribution of the position of votes across the four conditions: independent list (A), independent cloud (B), social influence list (C) and social influence cloud (D). A and C show the distribution of votes over order (0 is the at the top of the page, 60 is at the bottom), and B and D show heatmaps of user votes. All are colored such that white corresponds to the most engagement, and purple the least.}
 \label{fig:one}
\end{figure}

We start by assessing the engagement across conditions, which we measured using the total number of votes each participant cast.  We find that participants in worlds with the cloud design engaged with more ganimals than those in the worlds with the list design ($p<0.001$, posthoc, $p_{\text{fisher}}=0.$ --- the left side of Figure~\ref{fig:one}).  We also find a main effect of social influence, whereby participants in worlds with social influence engaged with more ganimals than those in the independent worlds ($p=0.0182$, posthoc, $p_{\text{fisher}}=0.0081$). 
We also look at the position of the votes for each of the conditions, as shown on the right side of Figure~\ref{fig:one}. For both the independent and social influence list views, we see a power law which closely matches the availability plot (Fig 2) from \citet{krumme2012quantifying} and the position bias plot (Fig 7) from \citet{abeliuk2017taming}. For the independent cloud, we see a constellation of white and yellow dots scattered uniformly across the cloud view. This stands in contrast to the social influence cloud, where we observe clustering near the center, with satellites around the periphery. The circle packing algorithm placed most top voted ganimals near the center, (e.g. see the social influence cloud in Figure~\ref{fig:exde}). This explains why worlds in the social influence cloud condition had the highest Gini coefficient on average: social influence induced users to click on top voted ganimals near the center, further increasing inequality. 
\begin{table*}
\centering
  \caption{Linear regression predicting average number of votes for the four conditions, with robust standard errors clustered on world.}
  \label{tab:engage}
  \begin{tabular}{lcccl}
    \toprule
     &Estimate & Standard Error & t value & p value \\
    \midrule
    {Social Influence} &   0.738& 0.271 &2.72&0.0182  \\
    {Cloud} & 1.322& 0.2608& 5.084&<0.001 \\
    {Intercept} & 3.208&  0.202& 15.860&<0.001 \\
    \midrule
          {$r^2=0.0427 $, N=549} &  & & & \\
    \bottomrule
  \end{tabular}
\end{table*}

\section{Discussion}
For both inequality and unpredictability, we observed the same pattern observed in the Music Lab, that social influence can increase the inequality and unpredictability of the success of content. The pattern was the same across the list view conditions which mirrored the original \rnr{layout}, and the cloud view conditions, which had a \rnr{different} layout. This suggests that the role of social influence in increasing inequality and unpredictability is separable from item position, and generalizes well to different \rnr{layout} patterns, and contexts with image media. Indeed, the original two-stage process of the Music Lab may be less relevant to today's digital platforms, since now images are the primary content and the primary outcome of interest is engagement with those images. Thus, replicating the Music Lab results for engagement with images shows that they generalize to modern contexts. 

Our results suggest that without social influence, many worlds converged towards a singular set of ganimal features. This status quo contained features that conform with morphological conventions of quality  (e.g. ganimals with eyes, a head, and dog-like features). Social influence, however, led to the rapid evolution of local cultures that dramatically diverged from this status quo, and that were ultimately more diverse. Many studies interpret the unpredictability that social influence induces to be a negative externality. Yet we found that this unpredictability corresponds to exploring novel areas of the possibility space, and led to more diverse and divergent local cultures. 

These findings stand in contrast to recent results which suggest that social influence decreases diversity, and thus undermines the wisdom of the crowds \cite{lorenz2011social}. It's important to note that the context in \citet{lorenz2011social} involved objective  performance tasks with ground truth, such as geographical facts and crime statistics. In such objective contexts, divergence from the status quo will ``lead the herd astray'’ by skewing responses away from the crowd average.  But in contexts like ours, where the popularity of novel cultural objects is \fed{inherently} subjective, such divergence can actually be a boon.

The inclusion of the cloud view allowed us to isolate \rnr{one possible mechanism} of how social influence impacts divergence and diversity: the number of ganimals visible at a given time. Since the list view shows one or two ganimals at a time and requires scrolling to see more, social influence is an effective cue for identify ganimals to engage with. In contrast, cloud view shows a large number of ganimals at a given time. This may explain why cloud view worlds were less diverse than list view worlds, regardless of social influence, since ganimals with morphological features traditionally associated with popularity could easily grab user's attention. \rnr{Indeed, since the size of the ganimals is so different between the two displays (quite large for list view and quite small for cloud view), it is hard to compare the two directly. This is one reason we focused primarily on the list view only, and future work should compare layouts that hold both the image size and number of images displayed constant across conditions}.


Our work has implications for the design of social media platforms. While social influence can lead to a ``winner take all'' market where smaller actors suffer, it can also lead to the rapid evolution of unexplored, diverse trends. Designers of social media platforms should use social influence responsibly to foster more heterogeneous notions of quality.  In particular, designers must recognize when \rnr{divergence} is desirable (e.g. fashion or other cultural artifacts) and when it is not  (e.g. for objective facts) and design interfaces that are appropriate to that situation. For instance, in the case of \rnr{subjective culture and trends, presenting feedback on the proportion of similar people may help promote divergence, whereas in situations with objective facts, feedback such as histogram of other people's opinion might be more appropriate.} In addition, the cloud view represents a new paradigm for newsfeed design that may be appropriate in contexts where it is desirable to mitigate social influence or the availability bias (such as objective contexts, as in \citet{lorenz2011social}).

Our work also has implications for how animal morphology relates to engagement. It has been shown that people have learned preferences for animal morphology on an evolutionary timescale and during development \cite{miralles2019empathy, gol2018polymorphism, colleony2017human} and \rnr{indeed many of the most popular ganimals have morphological traits of common pets like cats and dogs (see panels C and D of Figure}~\ref{fig:exde}). \rnr{However, our world-level experimental design allows us to decouple the role of visual familiarity/innate preference and social influence in ganimal popularity. Our results suggest that people do indeed use social influence to inform their preferences for animal morphology (at least in the context of AI-generated hybrid animals) and that social influence can lead to the formation of divergent local preferences.} In a world where ``charismatic megafauna'' --- animals that play into those more conventional evolutionary and developmental preferences (like the Giant Panda) --- absorb much of the public attention and funds for conservation \cite{estren2018ethics,miralles2019empathy, nautil,colleony2017human}, this suggests that social influence may be a powerful mechanism to invigorate attention towards the conservation of animals that do not have such morphological features, like the Chinese Sturgeon \cite{zhou2020chinese} or the blobfish \cite{blobfish}. 

Our work has several important limitations. First, the small number of worlds (4 per condition = 16) makes conducting statistical inference at the world level challenging. Future work could provide more precise effect sizes with a larger number of worlds (e.g. 50 worlds per condition, as recommended by \citet{abeliuk2017taming}). One particular way of achieving this for web-based experiments that rely on virality for recruitment (as this and the original Music Lab were) is to start with a small, fixed number of worlds per condition, and dynamically branch new worlds (with the same fixed seed set) once a max cap of participants in existing worlds has been reached. This would also naturally allow for cohort blocking to account for variation in outcomes over the course of the experiment's run. Another methodological limitation was the fact that we did not collect demographics to preserve both the privacy and fun of the experience. This lack of user demographics and covariates prevented us from assessing the representativeness of our users, as well as any heterogeneous effects.

We believe that the approach introduced by the Music Lab --- randomization at the world level, each with its own independently evolving local ecology ---  offers platform designers a rigorous way to assess how design interventions not only affect individual behavior as in standard A/B testing, but also complex, collective behavior.  We hope this work demonstrates that this experimental paradigm generalizes to outcomes beyond inequality and unpredictability, and in turn can promote interventions to increase the diversity of information ecosystems. 


\section{Acknowledgements}
We thank Océane Boulais,  Aurélien Miralles, David Rand, Dean Eckles, Matt Salganik, Adam Bear, Anna Chung, Neil Gaikwad, Morgan Frank, Esteban Moro, Skylar Gordon, Josh Hirschfeld-Kroen, Micah Epstein,  Jack Muller, Dima Smirnov and Adam Bear for helpful feedback and comments. 

\bibliographystyle{ACM-Reference-Format}
\bibliography{main}

\end{document}